\documentclass{elsart3p}

\usepackage{graphicx}

\usepackage[figuresright]{rotating}

\usepackage{amssymb}

\newcommand{\proton}{\ensuremath{p}}
\newcommand{\neutron}{\ensuremath{n}}
\newcommand{\pg}{(\proton,\ensuremath{\gamma})}
\newcommand{\pn}{(\proton,\neutron)}

\newcommand{\sfs}{$S_{17}$}
\newcommand{\bes}{$^7$Be}
\newcommand{\lis}{$^7$Li}
\newcommand{\bei}{$^7$Be$^{4+}$}
\newcommand{\lii}{$^7$Li$^{3+}$}
\newcommand{\boe}{$^8$B}
\newcommand{\lie}{$^8$Li}
\newcommand{\bpg}{\bes\pg\boe}

\newcommand{\helion}{\ensuremath{^3}He}
\newcommand{\hefour}{\ensuremath{^4}He}
\newcommand{\onem}{\ensuremath{1^{-}}}
\newcommand{\onep}{\ensuremath{1^{+}}}
\newcommand{\twom}{\ensuremath{2^{-}}}
\newcommand{\twop}{\ensuremath{2^{+}}}

\newcommand{\threep}{\ensuremath{3^{+}}}
\newcommand{\ecm}{\ensuremath{E_{\mathrm{cm}}}}
\newcommand{\eex}{\ensuremath{E_{\mathrm{ex}}}}
\newcommand{\thetalab}{\ensuremath{\theta_{\mathrm{lab}}}}

\begin{document}

\begin{frontmatter}



\title{Low-lying non-normal parity states in \boe\ measured by proton elastic scattering
on \bes}


\author[CNS]{H.~Yamaguchi\corauthref{cor1}}
\ead{yamag@cns.s.u-tokyo.ac.jp}
\author[CNS]{Y.~Wakabayashi} 
\author[CNS]{S.~Kubono} 
\author[CNS]{G.~Amadio} 
\author[CNS]{H.~Fujikawa} 
\author[KU]{T.~Teranishi}
\author[UT]{A.~Saito}
\author[CAS]{J.J.~He}  
\author[RIKEN]{S.~Nishimura} 
\author[Rikkyo]{Y.~Togano} 
\author[Chungang]{Y.K.~Kwon} 
\author[CNS]{M.~Niikura} 
\author[Tohoku]{N.~Iwasa}   
\author[Tohoku]{K.~Inafuku} and  
\author[Viet]{L.H.~Khiem}  


\corauth[cor1]{Corresponding author.}


\address[CNS]{Center for Nuclear Study (CNS), University of Tokyo, RIKEN campus, 
  2-1 Hirosawa, Wako, Saitama 351-0198, Japan}

\address[KU]{Department of Physics, Kyushu University, 
  6-10-1 Hakozaki, Fukuoka 812-8581, Japan}

\address[UT]{  Department of Physics, Graduate School of Science, University
  of Tokyo 7-3-1 Hongo, Bunkyo-ku, Tokyo 113-0033, Japan}

\address[CAS]{Institute of Modern Physics, CAS, Nanchang Road 363, 730000 Lanzhou,
People's Republic of China}

\address[RIKEN]{ The Institute of Physical and Chemical Research (RIKEN), 
2-1 Hirosawa, Wako, Saitama 351-0198, Japan}

\address[Rikkyo]{Department of Physics, Rikkyo University, Tokyo 171-8501, Japan}

\address[Chungang]{Department of Physics, Chung-Ang University, Seoul 156-756, South Korea}

\address[Tohoku]{Department of Physics, Tohoku University, Aoba, Sendai, Miyagi 980-8578, Japan}

\address[Viet]{Institute of Physics and Electronics, Vietnam Academy 
    of Science and Technology, 8 Hoang Quoc Viet St., Nghia do, Hanoi, Vietnam}


\begin{abstract}
A new measurement of proton resonance scattering on \bes\ was performed up to the center-of-mass
energy of 6.7 MeV using the low-energy RI beam facility CRIB (CNS Radioactive Ion Beam separator) at
the Center for Nuclear Study of the University of Tokyo. 
The excitation function of \bes+\proton\ elastic scattering above
3.5 MeV was measured successfully for the first time, 
providing important information about 
the resonance structure of the \boe\ nucleus.
The resonances are related to the reaction rate of \bpg, 
which is the key reaction in solar \boe\ neutrino production.
Evidence for the presence of two negative parity states is presented. 
One of them is a \twom\ state observed as a broad $s$-wave resonance, the existence of which
had been questionable.
Its possible effects on the determination of the 
astrophysical S-factor of \bpg\  at solar energy 
are discussed.
The other state had not been observed in previous measurements,
and its $J^{\pi}$ was determined as \onem.

\end{abstract}

\end{frontmatter}


The astrophysical S-factor \sfs($E$) of the \bpg\ reaction is one of the most important parameters in the
standard solar model, because its value at the energy of the 
solar center is directly related to the flux of the \boe\
neutrino, which is the dominant component of the solar 
neutrinos detected by some of the major neutrino observatories on
earth \cite{Fukuda:01,Ahmad:02}. 
\sfs\ should be determined with a precision greater
 than about 5\%, in the energy region
below 300 keV, in order to test the solar model 
by comparing the theoretical prediction for the \boe\ neutrino flux
with the observations \cite{Adelberger:98}. 
For this reason, a number of experimental groups have put in
great efforts in that direction
\cite{Vaughn:70,Filippone:83,Hammache:98,Hammache:01,Strieder:01,Kikuchi:98,Iwasa:99,Davids:01,Schumann:03,Baby:03,Junghans:03,Cyburt:04}. 
The precision of the existing data, however, is still limited because of 
the very small cross section of the \bpg\
reaction in such a low-energy region.

To evaluate \sfs\ at low energies, one needs information about 
the nuclear structure of \boe, which has
been poorly known until recently. Only the lowest two excited states,
 at 0.77 and 2.32 MeV, 
were observed clearly in previous experiments \cite{TUNL:04}. 
A broad 2$^{-}$ resonance was
observed around 3 MeV  \cite{Goldberg:98}, however,
negative parity is non-normal for nuclei with a mass number of 8,
and the \twom\ state  was explained as a low-lying 2s state. 
 In another measurement \cite{Rogachev:01}, 
the broad state was not 
directly observed; nevertheless, 
the spectrum was considered consistent with the presence of the state, 
if it is located at 3.5 MeV with a
width of 4 MeV, or more.
Such a broad resonance may affect the \bpg\ reaction rate in the energy
region far below 1 MeV. 
Investigating why a 2s state appears at such low energy is also
interesting, and  
there are studies that predict the presence of the 2$^{-}$ state
in \boe\ or its mirror nucleus, \lie\
\cite{Knox:81,Knox:87,Hees:83,Hees:84,Grigorenko:96,Barker:00}. 
The \twom\ resonance is possibly related to 
the proton-halo structure  of the \boe\ nucleus \cite{Minamisono:92}.
Thus, we intended to study the 
resonance structure of \boe\, to evidently observe the 3.5 MeV
resonance reported in previous measurements, and explore 
the unknown energy region above 3.5 MeV.



The measurement was performed using CRIB (CNS Radioactive Ion Beam 
separator) at
the Center for Nuclear Study (CNS) of the University of Tokyo
\cite{Kubono:02,Yanagisawa:05}.
CRIB can produce RI beams with the in-flight method,
using primary heavy-ion beams from 
the AVF cyclotron of RIKEN. 
The primary beam used in this measurement was \lii\ of 8.76 MeV/u
at 100 pnA.
The RI-beam production target was pure hydrogen 
gas at 760 Torr and room temperature ($\sim$ 300 K),
enclosed in an 8-cm-long cell. 
A secondary \bes\ beam at 53.8 MeV was produced from \lis\ 
via \pn\ reaction in inverse kinematics. 
The typical intensity of the \bei\ beam was $3 \times 10^5$ 
particles per second at the target of resonance scattering. 
A Wien filter was used for purification of the secondary beam.
The beam purity (the number ratio of \bei\ to the total), before  
and after passing through the Wien filter,  was 56\%  and 100\%,
respectively.

\begin{figure}
\centerline{\includegraphics[scale=0.7]{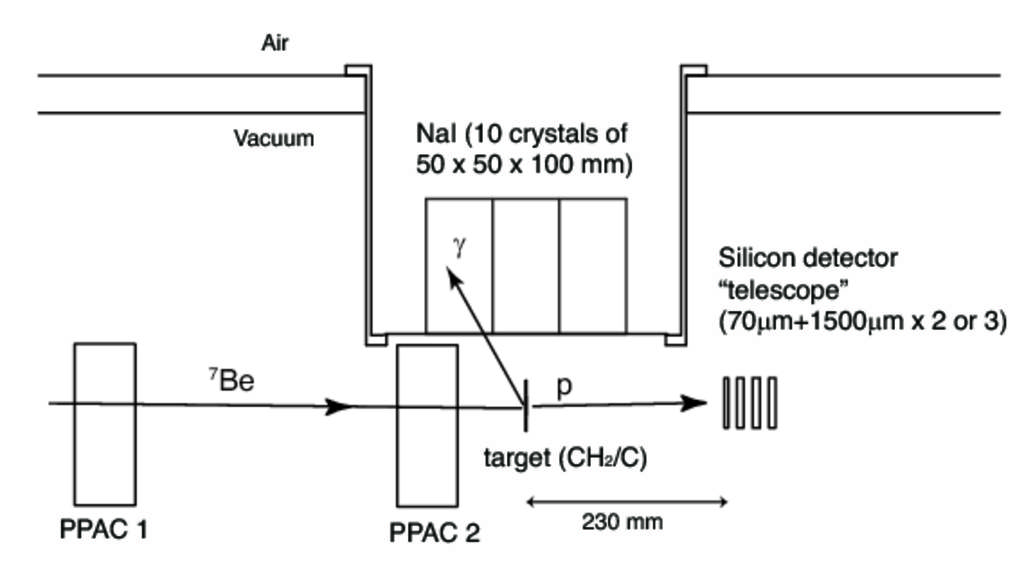}}
\caption{\label{fig:schematic} Arrangement of the detectors and targets 
in the experimental chamber.}
\end{figure}

We used an experimental method
similar to past measurements 
of proton elastic resonance scattering at CRIB \cite{Teranishi:03,Teranishi:07}.
A main feature of this method is a
thick target \cite{Artemov:90,Kubono:01}, 
which enables simultaneous measurements
of cross section of various excitation energies.
The targets and detectors for the scattering experiment were
in a vacuum chamber located at the end of the beam line.
Figure~\ref{fig:schematic} shows a schematic view 
 of  the  experimental setup
in the chamber.
Two parallel-plate avalanche counters (PPACs) \cite{Kumagai:01}
measured timing and position of the incoming \bes\ beam
with a position resolution of 1 mm or better. 
The timing signal was used for 
producing event triggers, and 
for particle identification 
using the time-of-flight (TOF) method.
The position and incident angle of the beam at the target were
determined by extrapolating the positions measured by PPACs.
The targets were films of 39-mg/cm$^2$-thick  polyethylene,  
and 54-mg/cm$^2$-thick carbon, both 
sufficiently thick to stop the \bes\ beam.
Carbon targets were used for evaluating background events originating
from carbon nuclei contained in the polyethylene target.
We accumulated data for 51 h with 
the polyethylene, and 17 h with the carbon target. 
Multi-layered silicon detector sets, referred to as $ \Delta$E-E telescopes, 
measured the energy and angular distributions of the recoiling protons.
Four telescopes were placed at a distance of 23 cm from the target to
cover the scattering angle in the laboratory frame \thetalab\
from 0 to 45 degrees.
Each telescope consisted of a 
thin ``$\Delta$E'' counter and two or three thick ``E'' counters,
each with an area of 50 $\times$ 50 mm.
The $\Delta$E counters were 60 to 75-$\mu$m thick, and 
divided into 16 strips for each side.
The 1.5-mm-thick E counters were 
placed behind the $\Delta$E counters.
NaI detectors were used for measuring 429-keV gamma rays
from inelastic scatterings to the first excited state of \bes.
We used ten NaI crystals, each with
a geometry of 50 $\times$ 50 $\times$ 100 mm,
covering 20\% of the total solid angle altogether.



Proton events 
were selected 
using measured energy ($\Delta E$-$E$) and timing information.
The center-of-mass energy 
\ecm\ of each event was 
determined from the measured proton energy and angle 
by calculations of kinematics and 
the energy loss in the target. 
Cross sections of the proton scattering events
for both the polyethylene and carbon targets  
were calculated from the 
number of proton events and
irradiated beam particles, 
the solid angle of the detector, and the target 
thicknesses.
The excitation function for the proton target was deduced 
by subtracting the carbon contribution
from the polyethylene spectrum. 

$E_{\mathrm{cm}}$ resolution of the excitation function 
was 40--70 keV in full width at half maximum (FWHM) at the most forward angle.
The uncertainty was mostly from 
energy straggling of the particles
in the thick target, 
along with the energy resolution of the silicon detectors.
At larger angles, the angular resolution of the recoiling proton 
produced large energy uncertainty 
and the resulting energy resolution was 
70--300 keV at $\thetalab=25$ degrees.

When the compound \boe\ nucleus has an excitation
 energy exceeding the threshold
at 1.72 MeV, decay to the 3-body channel (\hefour\ + \helion\ + \proton) may occur.
Background proton events from this 3-body-channel decay
distributed over wide energy and angular ranges
must be subtracted from the obtained excitation functions.
The energy and angular distributions of the background protons 
were estimated by a Monte Carlo simulation, assuming isotropic 
particle emissions in the center-of-mass frame.
The absolute value of the contribution was normalized by 
measured numbers of multiple hit (proton with  \hefour\ and/or \helion) events.
The estimated 3-body background contribution
was 30 mb/sr at maximum and structureless
in the excitation functions, and 
thus it is not very influential on the line shape.

We measured de-excitation $\gamma$ rays in
the inelastic events with the NaI detectors. 
These detectors had an energy resolution 
of 10\% (FWHM) for 662-keV gamma rays.
The 429-keV photopeak detection efficiency $\epsilon$ was 
measured as 7.1\%, using $\gamma$-ray sources
placed at the target position.
The $\gamma$-ray energy spectrum of 
proton-$\gamma$ coincident events showed an intense peak at 429 keV
and the contribution to the excitation function by the inelastic scattering 
events was successfully deduced.
The inelastic contribution, about 10\%
of the elastic scattering, was subtracted from the 
total excitation function.  

\begin{figure}
\centerline{\includegraphics[scale=0.9]{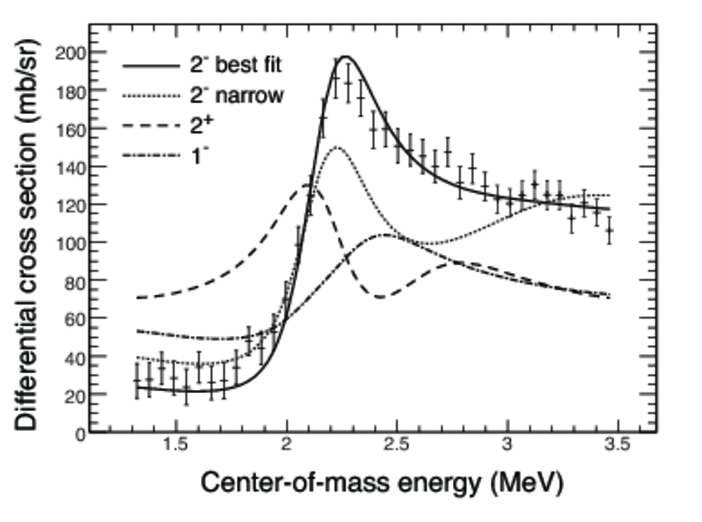}}
\caption{\label{fig:f_2-_fit_summary}  
Excitation function of \proton+\bes\ elastic 
scattering below 3.5 MeV, measured between 
0 and 8 degrees. R-matrix fit results,
with 
a broad \twom, \onem\ or \twop\ state, are also drawn. 
The dotted curve (\twom\ narrow) is the result with the \twom\ resonance 
having a width about half of the one in the best fit.}
\end{figure}

The presence of the \twom\ state around 3.5 MeV
was questionable because of the limited statistics 
and energy range in previous experiments
\cite{Goldberg:98,Rogachev:01}, although 
the \twom\ state was also expected to exist
from the analysis on the experimental data 
of mirror nucleus \cite{Lane:64,Knox:81,Knox:87,Grigorenko:96,Barker:00}
and shell model calculations \cite{Aswad:73,Hees:83,Hees:84}. 
We successfully performed measurements
with more counting statistics and a wider energy range,
and a slowly varying excitation function after the 
peak around 2.3 MeV was observed, as shown in Fig.~\ref{fig:f_2-_fit_summary}.
This excitation function strongly suggests that 
the peak at 2.3 MeV was enhanced by a broad state that 
locates at higher energy.
Following this assumption, which is virtually the same as 
the one taken in \cite{Rogachev:01}, 
we performed an R-matrix analysis, using SAMMY \cite{SAMMY:00} code.
The channel radius was fixed at 4.3 fm, 
the same value as in \cite{Rogachev:01} and \cite{Knox:81}.  
We confirmed that the result was not very sensitive to a 
deviation of channel radius within 0.5 fm.
Two known resonances at excitation energies 
\eex = 0.77 and 2.32 MeV were introduced in the fit
using parameters in \cite{TUNL:04},
although the former was not effective in our energy range.
The R-matrix calculations provide a reliable determination of the 
resonance parameters (energy $E$, width $\Gamma$, spin $J$, and parity $\pi$) 
even for such broad states.
In the best fit for $J^{\pi}=2^{-}$,
shown in Fig.~\ref{fig:f_2-_fit_summary},
\eex= 3.2 MeV and $\Gamma =3.8$ MeV. 
Although the \twom\ resonance is broad and 
did not appear as a distinct peak,
the excitation function was sensitive to 
variations of $E$ and $\Gamma$.
When we reduced $\Gamma$ by about half ($\Gamma$=1.8 MeV),
the resulting excitation function,
indicated as ``\twom\ narrow'' in the figure,
was in complete disagreement with the original function,
proving that $\Gamma$ of the \twom\ resonance truly affects
the calculated excitation function.
The resonance is considered to be an $s$-wave 
resonance, as it has negative parity and broad width.
We could not obtain satisfactory fits by introducing 
broad \onem\ or any possible positive parity state,
while a broad \twop\ state was 
introduced to reproduce the excitation function 
in a previous study \cite{Halderson:04}.

\begin{figure}
\centerline{\includegraphics[scale=1.0]{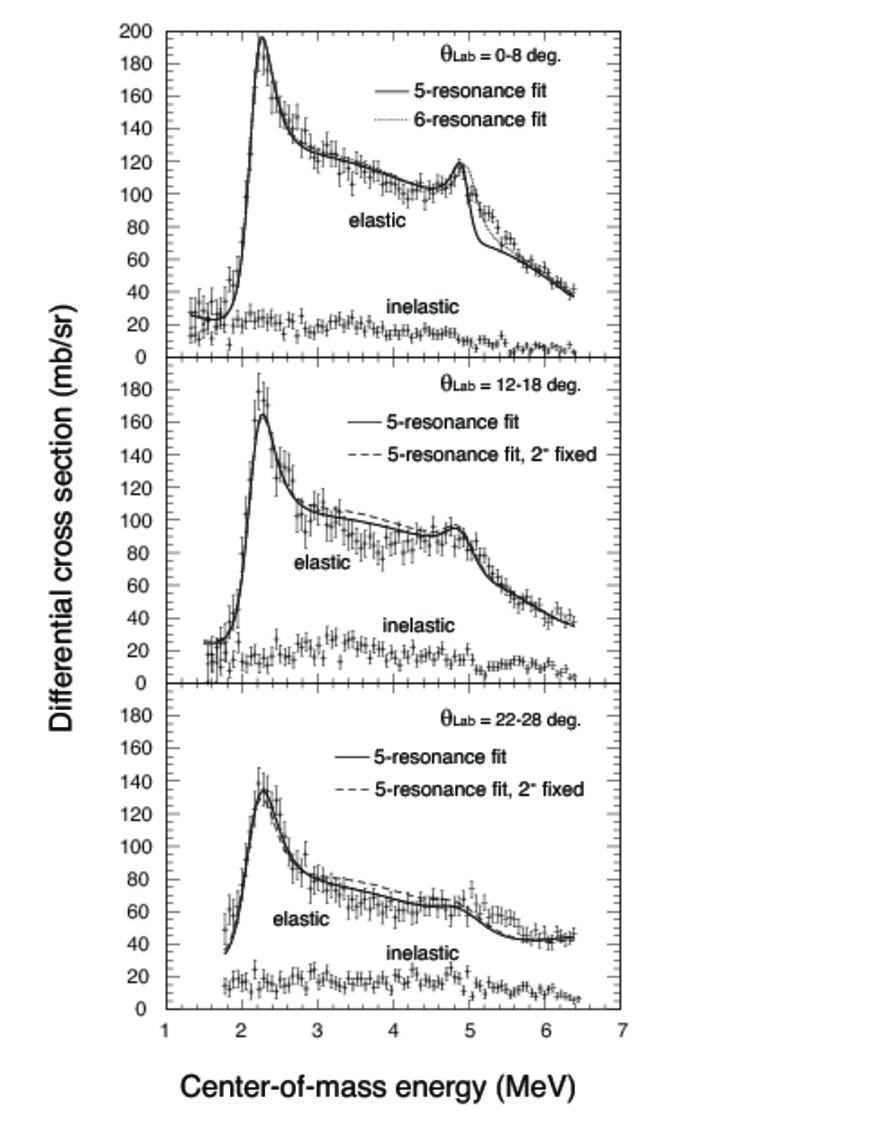}}
\caption{\label{fig:excitation_functions_fit}  
Excitation functions of \proton+\bes\ elastic scattering for three angular ranges,
fitted with R-matrix calculations.
Contributions from inelastic scattering are also shown.
The best fit for each angular range 
with five resonances, including two 
unknown resonances (\onem\ at 5.0 MeV and \threep\ at around 7 MeV) 
are shown with solid curves. 
The dashed curves for the larger two angular ranges are
the calculated functions
using the same $\twom$ resonance parameters 
as that between 0 and 8 degrees.
The dotted curve for 0--8 degrees is 
a 6-resonance fit with an additional \onep\ state at 5.8 MeV.
}
\end{figure}

Assuming the presence of the \twom\ state,
we expanded the R-matrix fit to the higher energy region,
as illustrated 
in Fig.~\ref{fig:excitation_functions_fit}.
The contribution of the inelastic scattering 
to the first excited state in \bes\ is shown in the same figure.
A characteristic peak structure was found around
the excitation energy of 5 MeV.
The peak is considered to be due to a resonance
that was not observed in previous studies.
R-matrix fits were performed introducing resonance 
around 5 MeV with all possible 
combinations of $J^{\pi}$, and only
\onem\ resonance with $s$- or $d$-wave provided reasonable fits.
The tail shape in the excitation function between 5.5 and 6.5 MeV was 
well-reproduced by introducing a \threep\ state,
which is known to exist in the mirror nucleus. 
The calculated excitation functions 
that fitted to the experimental data for three angular ranges 
are shown as solid curves in the figure. 
The parameters for the \twom\ resonance are 
consistent for all the angular ranges
within the experimental resolution, as shown by the
dashed curves in higher two angular ranges,
obtained using the same parameters for the \twom\ resonance
as the lowest angular range.
The \onem\ resonance was not observed clearly in the
spectra for larger angles, because of the limited 
energy resolution.
The resonance parameters for the newly introduced \threep\ state,
which provide best fits 
were \eex= 6.8--7.5~MeV and $\Gamma=$2--4~MeV, 
depending on the angular range.
The fit function shows small, but systematic 
deviations from the measured data, 
as seen around 5.5 MeV for the lowest angular range.
This may suggest that the excitation function 
cannot be reproduced by the sets of resonances
we assumed.
For example, the fit was improved by introducing another \onep\ state 
at 5.8 MeV, as shown with a dotted curve in the figure.
  
The resonance parameters
determined by the present work and previous studies
are summarized in Table~\ref{tab:resonance_params}.
\begin{center}
\begin{table}[htbp]
\caption{\label{tab:resonance_params} Resonance parameters of \boe\
determined by the present work and previous studies.
$l$ is the angular momentum used in the R-matrix calculation.}
\begin{tabular}{ccccc}
\hline\hline
$J^{\pi}$  & $l$ &\eex\ (MeV) & $\Gamma$ (MeV) & Reference \\
\hline   
\onep\    & 1 &0.7695 $\pm$ 0.0025 & 0.0356 $\pm$ 0.0006 & \cite{TUNL:04}\\
\hline   
\threep\ & 1 &2.32 $\pm$ 0.02 & 0.35 $\pm$ 0.03 & \cite{TUNL:04}\\
\hline
\twom\         & 0  & 3.2$^{+0.3}_{-0.2}$ &3.4$^{+0.8}_{-0.5}$ &present\\
(\twom, \onem) & 0 &3 & 1--4 & \cite{Goldberg:98}  \\
\twom          & 0 & 3.5 $\pm$ 0.5 &8 $\pm$ 4 & \cite{Rogachev:01} \\
\hline
1$^{-}$ &0 or 2& 5.0 $\pm$ 0.4 &0.15 $\pm$ 0.10 &present  \\
\hline
(3$^{+}$) & 1 & $\sim$7 & $>$2 &present \\
\hline\hline
\end{tabular}
\end{table}
\end{center}
The parameters for the \twom\ state were determined 
with improved precision, 
showing no large discrepancies with previous measurements.
Our excitation functions, including the 
angular dependence
and measurement of inelastic scattering, 
strongly support the existence of 
the broad 2$^{-}$ state in \boe\ nucleus around 3.2 MeV.
Excited states of \boe\ higher than 3.5 MeV were not explored
in past measurements, and we discovered new 
resonance at 5.0 MeV and assigned its $J^{\pi}$ as \onem.
A \onem\ resonance in the $A$=8 nuclei 
was predicted to emerge in the vicinity of 
a \twom\ state
by theoretical studies \cite{Aswad:73,Knox:87,Hees:83,Hees:84}. 
In \cite{Barker:00}, a structure due to \onem\ level
appeared at \eex\ = 4.1 MeV ($E_{\mathrm{proton}}$ = 4.5 MeV) 
in the calculated S-factor spectrum.
The observed resonance might be the first evidence 
for these predictions in \boe,
and could lead to extensive studies 
on the structure of the \boe\ nucleus. 
We found an indication of resonance at around 7~MeV,
but more evidence is required to determine its parameters.

\begin{figure}
\centerline{\includegraphics[scale=1.2]{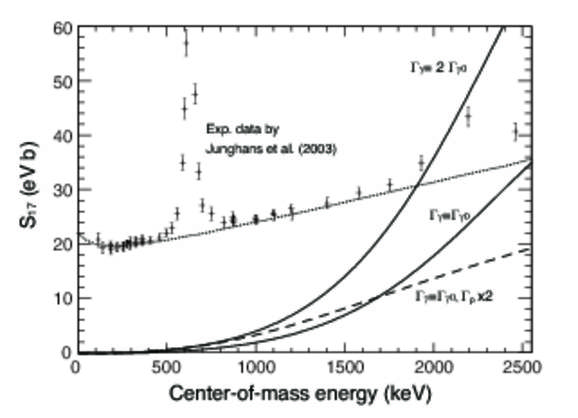}}
\caption{\label{fig:f_2-_sfactor}  
Resonant contributions of the \twom\ state to the astrophysical S-factor \sfs,
evaluated by the Breit-Wigner formula.
The experimental data and 
the nonresonant contribution (dotted curve) in \cite{Junghans:03} 
are also shown for comparison of the magnitude.
}
\end{figure}

In the precise determination of the \bes\pg\boe\ S-factor by
Junghans and coworkers 
\cite{Junghans:03}, the resonant contribution 
was evaluated by the Breit-Wigner function,
\begin{equation}
\sigma ( \ecm )=
\frac{C}{\ecm} 
\frac{\Gamma_p(\ecm) \Gamma_{\gamma}(\ecm)}{(\ecm - E_0)^2 + \Gamma_p(\ecm)^2 /4}. 
\label{sfac}
\end{equation}
The resonant contribution of the broad \twom\ resonance 
calculated  using our parameters and Eq.~\ref{sfac} is illustrated in Fig.~\ref{fig:f_2-_sfactor}. 
Because the gamma width $\Gamma_{\gamma}$ was not determined by 
our measurement, 
standard width $\Gamma_{\gamma 0}= 11$ eV, 
corresponding to the Weisskopf unit, was defined,
and the contributions were calculated 
for $\Gamma_{\gamma} = \Gamma_{\gamma 0}$ and 2$\Gamma_{\gamma 0}$
cases as shown in Fig.~\ref{fig:f_2-_sfactor}.  
The curve for $\Gamma_{\gamma} = \Gamma_{\gamma 0}$ shows a
considerable contribution, 
which may partly explain the structure of the 
the experimental data in the high energy (\ecm $\sim$ 2 MeV) 
region. 
$\Gamma_{\gamma} = 2 \Gamma_{\gamma 0}$ is an extreme case,
 and apparently such a contribution 
was not observed in the high energy region. 
Nevertheless, the contribution at the solar energy was negligible 
compared to the experimental precision. 
Even if $\Gamma_{p}$ was doubled, 
as shown by the dashed curve in the figure,
the resulting contribution at the solar energy was negligible.
Therefore, the resonant reaction by 
the \twom\ state is expected to be ineffective for 
the determination of \sfs.

In \cite{Junghans:03}, the nonresonant contribution was 
evaluated by calculations using a microscopic cluster model \cite{Descouvemont:94} 
and other methods \cite{Jennings:98}.
The model used in \cite{Descouvemont:94} implicitly 
involves the \twom\ state as the $s$-wave contribution,
but the contribution would not be very sensitive to 
the resonance parameters.
A realistic evaluation might be possible
by calculations that explicitly involve a \twom\ state,
such as the work by Barker and Mukhamedzhanov \cite{Barker:00,Barker:06}. 
They introduced a \twom\ level in \boe\ at \eex=3.0 MeV and $\Gamma=$3.7--5.2 MeV 
to explain the \lie+\neutron\ elastic scattering data.   
We obtained resonance parameters in agreement with these, and thus 
their discussion should not be altered significantly.

In summary,
we have studied the proton resonance scattering on \bes,
using a pure \bes\ beam produced at CRIB.
The excitation function of \boe\ was measured up to the excitation
energy of 6.7 MeV, using the thick-target method and
resonance parameters of two negative (non-normal) parity states were 
determined.
The \twom\ resonance at 3.2 MeV was reported in two previous measurements,
and we determined its energy and width with improved precision.
The effect of the \twom\ resonance 
on the determination of \sfs\ 
was estimated
to be small compared to the experimental precision. 
Another resonance at 5 MeV was observed for the first time,
and it is considered to be the \onem\ state predicted in 
theoretical studies.

The experiment was performed at RI Beam Factory operated
by RIKEN Nishina Center and CNS, the University of Tokyo.
We are grateful to the RIKEN accelerator staff for their help.
This work was supported by a  Grant-in-Aid for Young Scientists (B)
(Grant No. 17740135) of JSPS.

\bibliographystyle{elsart-num_mod}

\bibliography{crib}

\end{document}